\documentstyle[12pt,epsf]{article}
\begin{document}
\begin{center} {\large Ion-Conducting Polymers -- Quenched to Dynamic
Disorder} \\ 
Sarmishtha Mandal, S. Tarafdar$\footnote{email : sujata@juphys.ernet.in}$ \\
Condensed Matter Physics Research Centre,  Department Of Physics,  \\ 
Jadavpur University,  Calcutta  700032, INDIA. \\
Aninda Jiban Bhattacharyya$\footnote{email : aninda@boson.bose.res.in}$ \\ 
S.N. Bose National Centre For Basic Sciences  \\
JD Block, Salt Lake,  Calcutta  700091, INDIA. \end{center}
\begin{abstract} 
Ion conducting polymers have a biphasic character with crystalline 
as well as amorphous  phases. There is moreover, a dynamic disorder due to 
motion of polymer chain segments. The PEO-NH$_4$ClO$_4$ system undergoes a
crossover from a DLA-type morphology for low salt fraction $(X)$ to a structure 
 with polygonal spherulites. In the present communication  we show that the
low $X$ regime exhibits a variation of diffusivity with crystallinity 
typical of a quenched  system, whereas the high X regime has dynamic disorder 
with rapid rearrangement. 
\end{abstract} 
{\bf Pacs No : 66.10.Ed ; 66.30.-h, 61.41.+e; 61.43.Hv}
 \newpage 
\par Polymers such as PEO complexed with salts form good ionic conductors which
are used as electrolytes in different devices \cite{app}. These materials 
are biphasic with crystalline as well as amorphous phases present in a 
single sample. In addition there is continuous rearrangement of the polymer
chains, which aids in the ion transport process \cite{rat}, so the disorder
is said to be dynamic.
\par In a recent study of PEO-NH$_4$ClO$_4$ systems it was shown that there are
two distinct morphological regimes for varying  salt concentration and
the ion-conductivity can be related to the morphology \cite{{fr},{prb}}.
In the present communication we show that the two morphological regimes have
a striking difference with respect to the dynamic nature of disorder. For
low salt fraction $(X)$ at room temperature the system is essentially quenched, whereas for high
$X$ there is a rapid rearrangement initially, with a percolation-like
behaviour setting in later. As temperature increases the low $X$ regime 
also shows dynamic disorder. 
\par Bhattacharyya et al \cite{jpcm} carried out a computer simulation study of
diffusion on a square (2-dim) lattice with rearrangement. Here, a random 
walker explores a lattice consisting of a random distribution of two types of sites --- one good conducting 
and the other low conducting. $\chi$  is the fraction of crystalline (low conducting) sites. The
two types have different trapping times corresponding to amorphous (short
time -  $\tau_1$) and crystalline (long time - $\tau_2$) phases. The distribution
of sites changes after a characteristic time $\tau_r$, this incorporates
dynamic disorder into the model. An empirical relation for the diffusion
coefficient $D(\chi,\tau_1,\tau_2,\tau_r)$  
was suggested to be of the form 
\begin{equation}
4D = \frac{1}{\tau_{eff}(\chi,\tau_1,\tau_2,\tau_r)} = \frac{1}{\tau_{max}}. \frac
{\alpha_{2}}{\alpha_{1} + \alpha_{2}} + \frac{1}{\tau_{min}}.\frac {\alpha_{1}}
{\alpha_{1} + \alpha_2}
\end{equation}
where
\begin{equation} \tau_{max} = \tau_{1}(1 - \chi) + \tau_{2}\chi \hspace*{.5 cm} ; 
\hspace*{.5 cm} \frac{1}{\tau_{min}} = \frac{1 - \chi}{\tau_{1}} +\frac{\chi}{\tau_{2}}
\end{equation}
\begin{equation}
\alpha_{1} = 1 - exp \left (-\frac{\tau_{1}}{\chi \tau_{r}} \right )
\hspace*{.5 cm} ; \hspace*{.5 cm}
\alpha_{2}=exp \left (-\frac{\tau_{2}}{(1-\chi)\tau_{r}} \right )
\end{equation}
Exact results are obtained in the limits of small and large $\tau_r$.
 With $\tau_r 
\rightarrow \infty$ (a quenched system) $D$ is given exactly by the relation
\begin{equation}   D = \frac{1}{4\tau_{1}} \left [(1 - \chi) + \chi R \right ]^{-1}  \end{equation}
which  is a curve, and $\tau_r \rightarrow 0$ gives a linear relation
\begin{equation}   D = \frac{1}{4\tau_{1}} \left[ (1 - \chi) + \frac{\chi}{R} \right ]
  \end{equation} 
Here R = $\tau_2$/$\tau_1$. 
Figure 1 shows the variation of $D$ with the crystallinity ($\chi$) for 
different $\tau_r$ obtained in \cite{jpcm}. The analytical 
results obtained by Nitzan et al \cite{drug} on a random barrier model are
very similar.
\begin{figure}[h]
\centering
\epsfxsize=4.0in\epsfysize=3.0in
\epsfbox{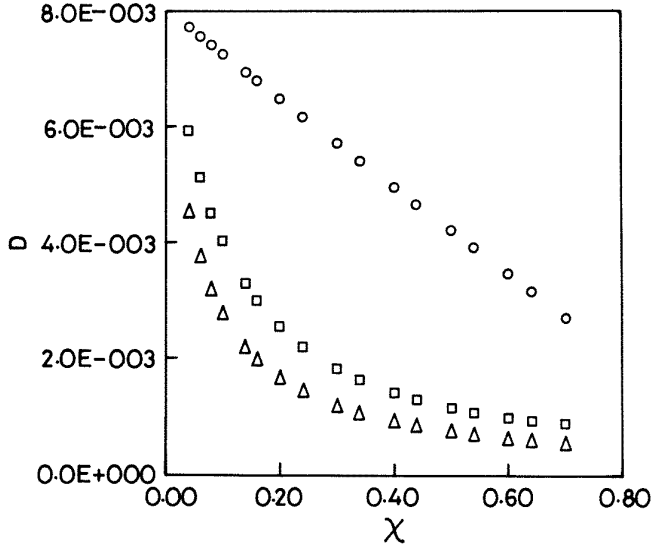}
\caption{ 
 Plot of diffusion coefficient  (D) versus crystallinity 
$\chi$ for different values of $\tau_r$ : ($\Box$)  for $\tau_r$ = 1.25$\times$ 10$^3$ using 
equation (1), ($\triangle$) for $\tau_r \rightarrow \infty$ using equation (4)
and  ($\circ$) for $\tau_r  \rightarrow 0 $ equation (5).} 
\label{Figure 1} 
\end{figure}
\par In the experimental work reported by Bhattacharyya et al \cite{prb}, ionic
conductivity ($\sigma$) and crystallinity ($\chi$) were both measured for
different salt fractions ($X$).
It was found that with $X$ increasing there is a rise in $\sigma$ accompanied by a
fall in $\chi$ upto $X_{tr}=0.18$. After this $\sigma$ falls and $\chi$
increases upto $X=0.30$, when the salt no longer goes into the complex.
\par The morphology for $X<X_{tr}$ is drastically different from $X>X_{tr}$
\cite{fr}. For low $X$, DLA type fractal clusters are formed in a uniform
background, whereas for high $X$ a polygonal pattern of spherulites is seen.
$\sigma$ is highest at the crossover region between the two different types
of morphology \cite{prb}.
\par In the present work we show that the low $X$ regime is almost quenched with
very large $\tau_r$, but for $X>X_{tr}$ there is dynamic disorder. To 
demonstrate this we first extract the diffusivity from the conductivity
data. This is done by assuming the Nernst-Einstein relation that
$$\sigma \propto D X$$
We assume here that $X$ is proportional to the number of charge carriers
present, so an effective diffusion coefficient $D'$ is obtained simply as
\begin{equation} D' = \sigma/X \end{equation} We are interested in the
variation of diffusivity with $\chi$, not the exact numerical value so the
variation in $D'$ is regarded as the variation in $D$. $D'$ vs $X$ and $\chi$
vs $X$ are shown in figure 2, for two different temperatures. It is clearly
apparent from the asymmetric nature of the curves about $X_{tr}$, that the
variation of $D'$ with $\chi$ on the left and right of $X_{tr}$ are 
different.
\begin{figure}[h] 
\centering
\epsfxsize=5.0in\epsfysize=3.0in
\epsfbox{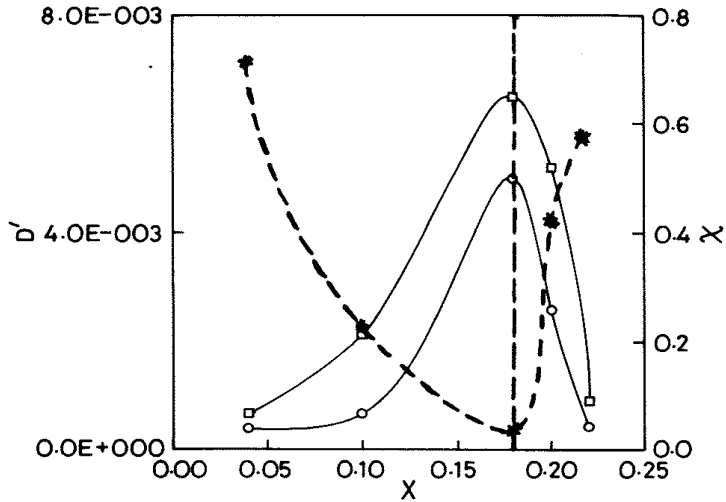}
\caption{Plot of D$'$ (continuous curves) at temperatures 
($\circ$) 35$^{\circ}$C and ($\Box$) 47$^{\circ}$C 
versus $X$. The dashed curve shows the plot of $\chi$ versus $X$ \cite{prb}.}
\label{Figure 2}
\end{figure}
\par From the curves in figure 2 we plot $D'$ vs $\chi$ for $X<X_{tr}$ and 
$X>X_{tr}$, the results are shown in figures 3 and 4. It is evident that for $X_<$
the curve resembles the $\tau_r \rightarrow \infty$ curve in figure 1,
whereas the $X_>$ curve is much closer to the $\tau_r \rightarrow 0$
curve.
\begin{figure}[p]
\centering
\epsfxsize=4.0in\epsfysize=3.0in
\epsfbox{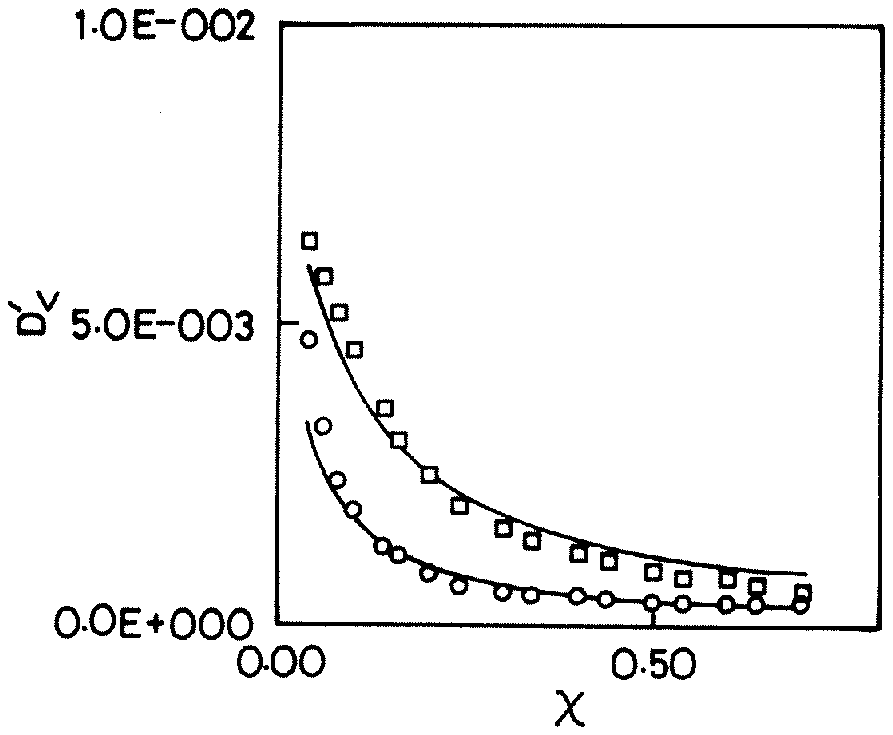}
\caption{
Plot of D$_<'$ at temperatures ($\circ$) 35$^{\circ}$C 
and  ($\Box$) 47$^{\circ}$C versus $\chi$. D$_<'$ (continuous lines) 
has been calculated using equation (1).}
\centering
\epsfxsize=4.0in\epsfysize=3.0in
\epsfbox{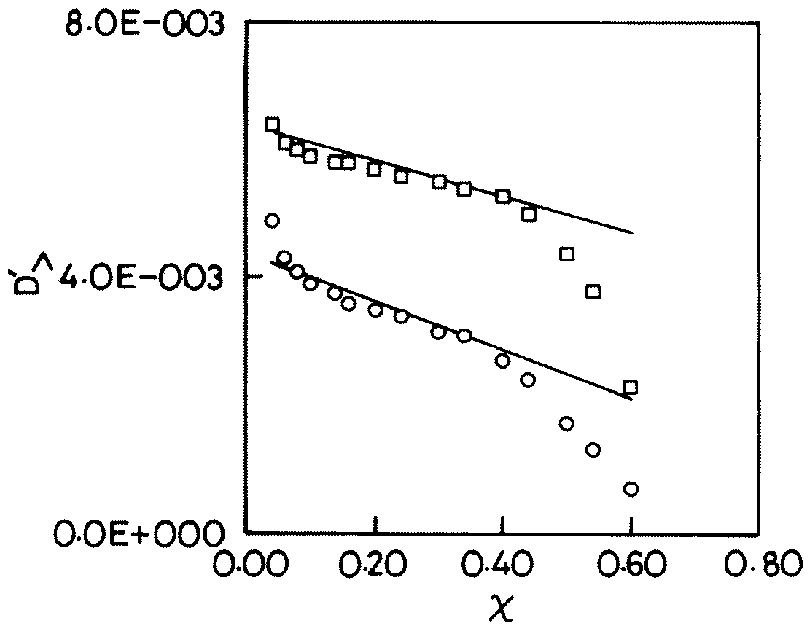}
\caption{
Plot of D$_>'$ at temperatures ($\circ$) 35$^{\circ}$C
 and  ($\Box$) 47$^{\circ}$C  versus $\chi$. D$_>'$ (continuous lines) 
has been calculated using equation (5).}
\end{figure}
\par Our results therefore indicate that for the low $X$ regime with the DLA type
morphology, the system is essentially quenched, but for the high $X$ regime
with the polygonal spherulites there is considerable dynamic disorder.
 As temperature increases both regimes show dynamic disorder. 
\par We estimate the relative values of $\tau_r$ and $\tau_1/\tau_2$ for the
two temperatures 35$^\circ$C and  47$^\circ$C by fitting the results for $D'$ to 
equation (5) in \cite{jpcm}.  
\par The best fit parameters are given in table 1 and the results are shown 
graphically in figures 3 and 4. Values of $\tau_1$, $\tau_2$ and $\tau_r$
are not in absolute units of time but relative values.
The extremely rapid rearrangement limit with linear
variation given by equation 5 is used to fit the $X_>$ regime. 
\vskip .3 cm \noindent
\begin{table}[h] 
\begin{center} 
Table 1 : Input parameters for calculation of D$'$.
\begin{tabular}{|c|c|c|c|c|c|c|} \hline 
& \multicolumn{3}{c|}{for $X_<$} & \multicolumn{3}{c|}{for $X_>$} \\
\cline{2-4} \cline{4-7} 
Temperature & $\tau_1$ & $\tau_2$ & $\tau_r$ & $\tau_1$ & $\tau_2$ & $\tau_r$ \\ 
\hline 
35$^{\circ}$C & 150 & 6000 & 2$\times$10$^4$ & 227.27 & 1666.66 & 0 \\ \hline
47$^{\circ}$C & 125 & 2500 & 5$\times$10$^3$ & 156.25 & 277.77  & 0  \\ \hline
\end{tabular} 
\end{center} 
\end{table}
\vskip .3 cm \noindent 
It may be
seen that the low $X$ regime fits the empirical relation for finite $\tau_r$
quite well, but for the high $X$ regime the experimental results deviate
linearly towards the point $D=0$ at $\chi=0.66$. The diffusivity vanishing
for a finite fraction of low conducting sites 0.66 is strongly reminiscent
of the results of percolation theory with one component totally insulating
\cite{perc}. In our model the crystalline fraction has a high but finite 
$\tau_2$, this gives a nonzero minimum of $D'$ at $\chi=1$.
\par The temperature variation of the parameters seen in Table I are physically 
realistic, with both $\tau_1$ and $\tau_2$ decreasing for increasing 
temperature indicating a higher mobility for both amorphous and crystalline 
fractions as temperature increases.
For the nearly quenched regime $\tau_r$
decreases as temperature increases, that is faster rearrangement occurs.
It may be mentioned that $log\sigma$ vs $1/T$ does not show linear Arrhenius
behaviour in our samples. This type of non-Arrhenius behaviour is often
observed in glassy or polymer samples \cite{narr}.
\par In conclusion we have shown that in PEO-ammonium perchlorate films two
distinct ion-conducting regions exist. One is a quenched regime with ion
conduction presumably taking place through the interstices of DLA type
fractal aggregates, and the other with polygonal spherulites where rapid
chain movement of the polymer assists in ion motion. It has been previously
suggested by Crist et al \cite{chris} that interlamellar  regions of spherulites
may be preferred for ion motion.
\par We are at present working on fluorescence spectroscopy studies of the polymer
samples to further substantiate our inferences.
\vskip .1 truecm \noindent {\bf Acknowledgement} : 
We thank the UGC for financial assistance. SM  thanks the CSIR for research fellowship.  

\end{document}